\documentclass[twocolumn,showpacs,preprintnumbers,amsmath]{revtex4}
\usepackage{amssymb}

\usepackage{graphicx}
\usepackage{dcolumn}
\usepackage{bm}

\begin{document}

\title{Jaynes-Cummings model: What emerges first beyond the rotating-wave
approximation?}
\author{Shu He$^{1,2}$, Qing-Hu Chen$^{1,3,*}$, Xue-Zao Ren$^{2}$, Tao Liu$^{2}$,
and Ke-Lin Wang$^{4}$}

\address{
$^{1}$ Department of Physics, Zhejiang University, Hangzhou 310027,
P. R. China \\
$^{2}$ School of Science,  Southwest University of  Science and
Technology, Mianyang 621010, P.  R.  China\\ $^{3}$ Center for
Statistical and Theoretical Condensed Matter
Physics, Zhejiang Normal University, Jinhua 321004, P. R. China  \\
$^{4}$ Department of Modern Physics, University of  Science and
Technology of China,  Hefei 230026, P.  R.  China
 }

 \date{\today}

\begin{abstract}
The Jaynes-Cummings model without the rotating-wave approximation
can be solved exactly by extended Swain's ansatz with the conserved
parity. The analytical approximations are then performed at
different levels. The well-known rotating-wave approximation is
naturally covered in the present zero and first approximations. The
effect of the counter rotating-wave term emerges clearly in the
second order approximation. The concise analytical expressions are
given explicitly and can be applicable up to the ultra-strong
coupling regime. The preliminary  application to the vacuum Rabi
splitting is shown to be very successful.
\end{abstract}

\maketitle

\section{introduction}

The interaction of light with matter is a fundamental one in optical
physics. The simplest paradigm is a two-level atom coupled to the
electromagnetic mode of a cavity. In the strong coupling regime
where the coupling strength $g/\omega $ ($\omega $ is the cavity
frequency) between the atom and the cavity mode exceeds the loss
rates, the atom and the cavity can repeatedly exchange excitations
before coherence is lost. The  Rabi oscillations can be observed in
this strong coupling  atom-cavity system, which is usually called as
cavity quantum electrodynamics (QED) \cite{CQED}. Typically, the
coupling strength in cavity QED reaches $g/\omega \thicksim
10^{-6}$. It can be described by the well-known Jaynes-Cummings (JC)
model\cite {JC}.

Recently, for superconducting qubits, a one-dimensional (1D)
transmission line resonator\cite{Wallraff} or a  LC
circuit\cite{Chiorescu,Wang,Deppe} can play a role of the cavity,
which is known today as circuit QED.  More recently, LC resonator
inductively coupled to a superconducting
qubit\cite{Niemczyk,exp,Mooij} has been realized experimentally. The
qubit-resonator coupling has been strengthened from $g/\omega
\thicksim 10^{-3}$ in the earlier realization\cite{Wallraff}, a few
percentage later \cite{Fink,Deppe},  to most recent ten
percentages\cite{Niemczyk,exp,Mooij}. In cavity QED system, the
rotating-wave approximation (RWA) is usually made, however, in the
circuit QED, due to the  ultra-strong coupling strength $g/\omega
\thicksim 0.1$ \cite{Niemczyk,exp,Mooij}, evidence for the breakdown
of the RWA has been provided\cite{Niemczyk}. Therefore,  counter
rotating-wave terms (CRTs) in the JC model  should be included.
Recently, many works have been devoted to this qubit-oscillator
system in the ultra-strong coupling
regime\cite{Werlang,Hanggi,Nori,Hwang,Hausinger,chen10b}.

Actually, the JC model without the RWA has been studied extensively
for more than 40 years. A incomplete list is given by Refs.
\cite{Swain,Kus,Durstt,Tur,Bishop,Feranchuk,Liu1,Irish,chenqh,liu,chen10,Pan,Casanova,QingHu1}.
Two main schemes are employed. One is based on the photonic Fock
states\cite{Swain,Kus,Durstt,Tur,Feranchuk,Bishop} with the pioneer
work  by Swain\cite{Swain}.  The  continued fractions are present
and the solution ten becomes very intricate. The other  is on the
basis of various polaron-like transformations or shifted operators,
which are basically photonic coherent states
approaches\cite{Liu1,Irish,chenqh,liu,chen10,Pan,Casanova,QingHu1}.
The very accurate solution can be obtained readily, but the infinite
photonic Fock states should be involved.

The RWA eigenstates only include two bare states, which have
facilitated  earlier clean investigations on  various quantum
phenomena in quantum optics. Surprisingly, one or a few dominant
terms in the eigenstates of the JC Hamiltonian beyond the RWA ones
are still lacking or not given explicitly until now, to the best of
our knowledge. What emerges first beyond the RWA results may be very
useful to analyze the effect of  CRTs on various phenomena at the
microscopic level. In this sense, a few dominant terms are more
helpful than the exact solution including infinite bare states.

In this paper, by using the conserved parity, we extend Swain's
wavefunction  to  the JC model without the RWA. We will not follow
the usual exact diagonalization routine. Alternatively, we derive a
polynomial equation with only single variable, just the eigenvalue
that we seek. The solutions to this polynomial equation can give
exactly all eigenfunctions and eigenvalues  for arbitrary
parameters. Moveover, we can perform  approximations step by step
with the help of these exact solutions. The zero and  first order
approximations will exactly recover the RWA results. The dominant
effect of the CRTs emerges in the second order approximation.

Without the RWA, the Hamiltonian of a two-level atom (qubit) with
transition frequency $\Delta $ interacting with a single-mode
quantized cavity of frequency $\omega $ is
\begin{equation}
H=\frac \Delta 2\sigma _z+\omega a^{\dagger }a+g\left( a^{\dagger }+a\right)
\sigma _x,
\end{equation}
where $g$ is coupling strength. $\sigma _x$ and $\sigma _z$ are
Pauli spin-$ 1/2$ operators, $a^{\dagger }$ and $a$ are the creation
and annihilation operators for the quantized field. Here, $\delta
=\Delta -\omega $ is defined as the dimensionless detuning
parameter. The energy scale is set $ \omega =1\;$here.

The RWA is made by neglecting the CRTs, $a^{\dagger }\sigma
_{+}+a\sigma _{-}$, then one can easily diagonalize the Hamiltonian
and obtain the eigenfunctions as\cite{ Scully}
\begin{equation}
\left| {}\right\rangle^{RWA}=\left( \
\begin{array}{l}
c_n\left| n\right\rangle \\
d_n\left| n+1\right\rangle
\end{array}
\right) ,n=0,1,2,...  \label{SRWA}
\end{equation}
For later use, we also list the relevant eigenvalues
\begin{eqnarray}
E_{1n}^{RWA} &=&n+\frac 12-\frac 12R_n,  \label{E1nRWA} \\
E_{2n}^{RWA} &=&n+\frac 12+\frac 12R_n,  \label{E2nRWA}
\end{eqnarray}
where $R_n=\sqrt{\delta ^2+4g^2\left( n+1\right) }$. In the ground
state (GS), the qubit is in  GS and the photon is in a vacuum state.
The GS energy is $E^0=-\frac \Delta 2$.

\begin{figure}[tbp]
\includegraphics[width=8cm]{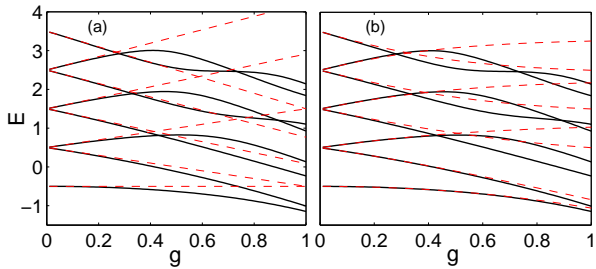}
\caption{ The first 8 energy levels as a function of coupling constant $g$
at resonance. The exact results without the RWA are presented by black solid
line. (a) The RWA results (red dashed line) and (b) The analytical results
in Eqs. (\ref{E1n}), (\ref{E2n}), (\ref{EGS}), and (\ref{E1EX}) of the
second-order approximation (red dashed line). }
\label{energy}
\end{figure}

Associated with JC Hamiltonian with and without the RWA is a
conserved parity $\Pi $, such that $\left[ H,\Pi \right] =0$, which
is given by
\begin{equation}
\Pi =\sigma _x\exp \left( i\pi \widehat{N}\right),
\end{equation}
where $\widehat{N}=$ $a^{+}a$ is the bosonic number operator. $\Pi $
has two eigenvalues $\pm 1$, depending on whether the excitation
number is even or odd. The above two states (6) and (7) with even
$n$ are of odd parity and with odd $n$  even parity. The ground
state is of even parity. The RWA results for the first $8$ energy
levels at resonance, $E_{1(2)n}=n+\frac 12 \pm g\sqrt{n+1)} $, are
given in Fig. \ref{energy} (a) for later comparisons.

\section{Exact solution without the RWA}

First we introduce a scheme to obtain the exact solutions to the JC
model without the RWA. For convenience, we can write a transformed
Hamiltonian with a rotation around an $y$ axis by an angle $\frac
\pi 2$. In the Matrix form it takes
\begin{equation}
H=\left(
\begin{array}{ll}
a^{\dagger }a+g(a^{\dagger }+a)-\alpha & -\frac \Delta 2 \\
-\frac \Delta 2 & a^{\dagger }a-g(a^{\dagger }+a)
\end{array}
\right).
\end{equation}

About 40 years ago, Swain first proposed an ansatz for the
wavefunction in the photonic Fock states\cite{Swain}, which is also
a starting point of the standard numerically exact diagonalization
(ED) scheme. Since then, various methods have been developed along
this line \cite{Kus,Durstt,Tur,Bishop}, but the conserved parity has
not been considered, to our knowledge. Therefore continued fractions
are unavoidably presented in the expressions for the eigensolutions.

We also proceed along this line, but the parity is incorporated in the
Swain's ansatz, which is given by
\begin{equation}
\left| {}\right\rangle =\left( \
\begin{array}{l}
\sum_{n=0}^Mc_n\left| n\right\rangle \\
\pm \sum_{n=0}^M(-1)^nc_n\left| n\right\rangle
\end{array}
\right)  \label{wavefunction}
\end{equation}
with $+(-)$ stands for even (odd) parity, M is the truncated number. The Schr$%
\stackrel{..}{o}$dinger equation gives
\begin{eqnarray}
\sum_{n=0}a^{\dagger }ac_n\left| n\right\rangle +g\sum_{n=0} &c_n&\left(
\sqrt{n}\left| n-1\right\rangle +\sqrt{n+1}\left| n+1\right\rangle \right)
\nonumber \\
\mp \frac \Delta 2\sum_{n=0}(-1)^nc_n\left| n\right\rangle
&=&Ec_n\left| n\right\rangle.
\end{eqnarray}
Left multiplying the photonic states $\left\langle m\right| $ gives
\[
mc_m+g\sqrt{m+1}c_{m+1}+g\sqrt{m}c_{m-1}\mp \frac \Delta
2(-1)^mc_m=Ec_m,
\]
then we have a recurrence relation
\begin{equation}
c_{m+1}=\frac 1{g\sqrt{m+1}}\left[ E-m\pm \frac \Delta
2(-1)^m\right] c_m- \sqrt{\frac m{m+1}}c_{m-1}.  \label{recurrence}
\end{equation}
By careful inspection of Eq. (\ref{wavefunction}), one can find that $c_0$
is flexible in the Schr$\stackrel{..}{o}$dinger equation where the
normalization for the eigenfunction is not necessary, so we select $c_0=1.0$%
. Then we have
\[
c_1=\frac 1g\left[ E\pm \frac \Delta 2\right].
\]
Once the first two terms are fixed, the coefficients of the other terms
higher than $a^{+}$ should be determined by the recurrence relation Eq. (\ref
{recurrence})

For $m=M$, the terms higher then $\left( a^{+}\right)^M$ are neglected, we
may set $c_{M+1}=0$, then we have
\begin{equation}
\left[ E-M\pm \frac \Delta 2(-1)^M\right] c_M-g\sqrt{M}c_{M-1}=0.
\label{centralEQ}
\end{equation}
Note that this is actually a one-variable polynomial equation of
degree $M$. The variable is just the eigenvalue we want to obtain.
It is expected that the roots of Eq. (\ref{centralEQ}) would give
the exact solutions to the JC model without the RWA if M is large
enough.

To obtain the true exact results, in principle, the truncated number
$M$ should be taken to infinity. Fortunately, it is not necessary.
It is found that finite terms in state (\ref{wavefunction}) are
sufficient to give exact results in the whole coupling range. The
usual numerical exact diagonalization can readily give the energy
levels and their wavefunctions in the JC model, which can be
regarded as a benchmark. We will increase the truncated number $M$
until the relative difference of the energies obtained from the
roots and the standard value is just less than $10^{-7}$, which sets
the criterion for the convergence achieved. Interestingly, for
coupling constant $g\le 0.1$ for three typical atom frequency
$\Delta =0.1,1$, and $1.5$, the truncated number $M=15$ in the
polynomial equation can give exactly $20$ energy levels by the above
criterion for convergence. For $g=1.0$ and $2.0$, the polynomial
equations with $M=25$ and $40$  can yield   about $ 20$ energy
levels exactly. In fact, all above calculations can be immediately
done in an ordinary PC.

For late use, the first 8 exact energy levels as a function of the
coupling constant $g$ obtained by the above polynomial equations
with $M=25$ at resonate case ($\Delta =1$) are presented in Fig.
1(a) and (b) by black solid lines. The parity is not changed after
the level crossing.

We then try to follow the energy curves to get some analytical
approximate results in the next section.

\section{Analytical results without RWA}

The recurrence relation Eq. (\ref{recurrence}) can be simplified to
a tridiagonal form
\[
\left(
\begin{array}{lllll}
\Omega _0(E) & 1 & 0 & ... & 0 \\
1 & \Omega _1(E) & \sqrt{2} & ... & 0 \\
0 & \sqrt{2} & \Omega _2(E) & ... & 0 \\
\ ... & \ ... & ... & ... & ... \\
0 & 0 & 0 & \sqrt{M} & \Omega _M(E)
\end{array}
\right) \left(
\begin{array}{l}
c_0 \\
c_1 \\
c_2 \\
... \\
c_M
\end{array}
\right) =0,
\]
where
\begin{equation}
\Omega _m(E)=\frac 1g\left[ m-E\mp \frac \Delta 2(-1)^m\right].
\label{kernel}
\end{equation}
The eigensolutions can be obtained from the zeros of the determinant
of the Matrix,

The zero-order approximation $\Omega _m(E)=0\;$gives
\[
E^m=m\mp (-1)^m\frac \Delta 2.
\]
The lowest energy is
\[
E^0=-\frac \Delta 2.
\]
The ground-state is of even parity, so the state is
\begin{equation}
\left| 0\right\rangle^{^{\prime }} =\left( \
\begin{array}{l}
c_0\left| 0\right\rangle \\
c_0\left| 0\right\rangle
\end{array}
\right).
\end{equation}
Transforming back to the original frame gives
\[
\left| 0\right\rangle =\frac 1{\sqrt{2}}\left(
\begin{array}{ll}
1 & -1 \\
1 & 1
\end{array}
\right) \left( \
\begin{array}{l}
c_0\left| 0\right\rangle \\
c_0\left| 0\right\rangle
\end{array}
\right) \varpropto \left| g,0\right\rangle,
\]
Interestingly, the first element in the zero-order approximation
actually gives exactly the ground-state in the RWA. The other
solutions are $g$ irrelevant and therefore  omitted.

\subsection{The first approximation}

The first approximation is made by selecting the matrices with two
order along the diagonal line
\begin{equation}
\left(
\begin{array}{ll}
\Omega _m(E) & \sqrt{m+1} \\
\sqrt{m+1} & \Omega _m(E)
\end{array}
\right) \left(
\begin{array}{l}
c_m \\
c_{m+1}
\end{array}
\right) =0.  \label{det1st}
\end{equation}
It is expected that the $m-th$ two order determinant would contain
the information of the two levels with same parity. Comparing with
the RWA results, it can be inferred that even $m$ is corresponding
to odd parity and odd $m$ \ to the even parity. Fortunately, we have
the same $\Omega (E)$ for any value of $m$
\[
\Omega _m(E)=\frac 1g\left[ m-E+\frac \Delta 2\right].
\]
Then we have following quadratic equation
\[
\left( m-E+\frac \Delta 2\right) \left( m+1-E-\frac \Delta 2\right) -\left(
m+1\right) g^2=0,
\]
which yields the eigenvalues
\begin{equation}
E=m+\frac 12\pm \frac 12\sqrt{\left( 1-\Delta \right) ^2+4g^2\left(
m+1\right)}.  \label{E1st}
\end{equation}
They are just the RWA ones in Eqs. (\ref{E1nRWA}) and (\ref{E2nRWA}) .

According to the wavefunction Eq. (\ref{wavefunction}), the eigenstate is
then obtained as
\begin{equation}
\left| m\right\rangle ^{\prime }=\left( \
\begin{array}{l}
c_m\left| m\right\rangle +c_{m+1}\left| m+1\right\rangle \\
\pm (-1)^m\left[ c_m\left| m\right\rangle -c_{m+1}\left| m+1\right\rangle
\right]
\end{array}
\right).
\end{equation}
By the above relation between $m\;$and parity, we always have $\pm (-1)^m=-1.
$ Transforming back to the original frame gives
\begin{equation}
\left| m\right\rangle =\frac 1{\sqrt{2}}\left(
\begin{array}{ll}
1 & -1 \\
1 & 1
\end{array}
\right) \left| m\right\rangle ^{\prime }\varpropto \left( \
\begin{array}{l}
c_m\left| m\right\rangle \\
c_{m+1}\left| m+1\right\rangle
\end{array}
\right),
\end{equation}
which are just the eigenstates under the RWA in Eq. (\ref{SRWA}) for
excited states.

So in the first approximation, we can not obtained results more than
the RWA ones for all excited states. The effect of the CRTS should
only emerges beyond the first approximation.

\subsection{The second order approximation}

Naturally, the second order approximation is performed by reducing
to the $ m-th$ three order determinant as
\begin{equation}
\left|
\begin{array}{lll}
\Omega _m(E) & \sqrt{m+1} & 0 \\
\sqrt{m+1} & \Omega _{m+1}(E) & \sqrt{m+2} \\
0 & \sqrt{m+2} & \Omega _{m+2}(E)
\end{array}
\right| =0.  \label{det2nd}
\end{equation}
For more concise, we only consider the resonant case $\delta =0$. It
is straightforward to extend to finite detunings.

Two univariate cubic equations for even and odd parity can be
explicitly derived for any $m-th$ three order determinant. Three
roots $y_1<y_2<y_3$ for each univariate cubic equation can be
obtained easily by the formula presented in the Appendix A.
Comparing with the exact ones obtained above, one can find that the
second roots $y_2$ are the true solutions. Especially, the first
root $y_1$ for $m=0$ with even (odd) parity gives the energy in the
GS (the first excited state). With these results in mind, the
general solutions can be grouped as the following two cases, and all
eigenvalues and eigenfunctions can be given analytically in the
unified way.

\subsubsection{$m=2k$ with even parity and $m=2k+1$ with odd parity}

For both $m=2k$ with even parity and $m=2k+1$ with odd parity
($k=0,1,2,...$
), we have the same univariate cubic equation in the following form $%
\allowbreak $
\[
E^3-\left( 3m+\frac 52\right) E^2
\]
\[
+\left[ \left( m+\frac 32\right) \left( 3m+\frac 12\right) -\left(
2m+3\right) g^2\right] E
\]
\begin{equation}
-\allowbreak \left( m-\frac 12\right) \left( m+\frac 32\right)
^2+\left[ 2m^2+4m+\frac 12\right] g^2=0.  \label{case1}
\end{equation}
According to the Appendix A, we have
\begin{eqnarray*}
A &=&4+\allowbreak \left( 6m+9\right) g^2, \\
B &=&-8m-12-\allowbreak \left( 12m^2+22m-3\right) g^2, \\
C &=&\allowbreak \left( 2m+3\right) ^2g^4+\left( 6m^3+13m^2+\frac 32m-\frac
34\right) g^2 \\
&&+\allowbreak 4m^2+12m+9.
\end{eqnarray*}
It can be readily proven $\Gamma =B^2-4AC<0\;$in this case, so there
are three different real roots. Note above that the energy level is
given by the second root $y_2$, so
\begin{equation}
E=\frac{\left( 3m+\frac 52\right) +\sqrt{4+\allowbreak \left(
6m+9\right) g^2 }\left[ \cos \theta -\sqrt{3}\sin \theta \right] }3,\label{E1nn}
\end{equation}
with
\[
\theta =\frac 13\arccos \left( \frac{\allowbreak 8-\left( 9m+27\right) g^2}{%
\sqrt{\left[ 4+\allowbreak \left( 6m+9\right) g^2\right]
^3}}\right).
\]
Especially, $m=0$ with even parity will give the GS additionally. The GS
energy is given by the first root $y_1$
\begin{equation}
E_{GS}=\frac{\frac 52+\sqrt{4+\allowbreak 9g^2}\left[ \cos \theta +\sqrt{3}%
\sin \theta \right] }3,
\end{equation}
with
\begin{equation}
\theta =\frac 13\arccos \left( \frac{\allowbreak
8-27g^2}{\sqrt{\left( 4+\allowbreak 9g^2\right) ^3}}\right).
\nonumber
\end{equation}
The states in this case all takes the form
\begin{equation}
\left| {}\right\rangle \varpropto \left( \
\begin{array}{l}
c_{m+1}\left| m+1\right\rangle \\
c_m\left| m\right\rangle +c_{m+1}\left| m+2\right\rangle
\end{array}
\right).
\end{equation}

\subsubsection{$m=2k+1$ with even parity and $m=2k\ $ with odd parity}

For both $m=2k+1$ with even parity and $m=2k\ $ with odd parity, we have the
same univariate cubic equation in the other form$\allowbreak $%
\[
E^3-\left( 3m+\frac 72\right) E^2
\]
\[
+\left[ \left( m+\frac 12\right) \left( 3m+\frac{11}2\right) -\left(
2m+3\right) g^2\right] E
\]
\begin{equation}
-\left( m+\frac 12\right) ^2\left( m+\frac 52\right) +\left(
2m^2+6m+\frac 72\right) g^2=0.  \label{case2}
\end{equation}
Similarly, we have
\begin{eqnarray*}
A &=&\left( 6m+9\right) g^2+4, \\
B &=&\allowbreak \left( -12m^2-38m-21\right) g^2+\allowbreak \left(
-8m-4\right), \\
C &=&\allowbreak \left( -2m-3\right) ^2g^4+\left( 6m^3+29m^2+\frac{83}2m+%
\frac{81}4\right) g^2 \\
&&+4m^2+4m+1.
\end{eqnarray*}
One can also readily prove $\Gamma =B^2-4AC<0$, So there are also
three different real roots. The energy level is given by the second
root$\;y_2$,
\begin{equation}
E=\frac{\left( 3m+\frac 72\right) +\sqrt{\left( 6m+9\right)
g^2+4}\left[ \cos \theta -\sqrt{3}\sin \theta \right] }3, \label
{E2nn}
\end{equation}
with
\[
\theta =\frac 13\arccos \left( \frac{-8+9mg^2}{\sqrt{\left[ 4+\left(
6m+9\right) g^2\right] ^3}}\right).
\]

Especially, $m=0$ with odd parity will yield the first excited state
additionally. The corresponding eigenenergy is given by the first root $y_1$
\begin{equation}
E_1=y_1=\frac{\frac 72+\sqrt{9g^2+4}\left[ \cos \theta +\sqrt{3}\sin
\theta \right] }3,
\end{equation}
with
\[
\theta =\frac 13\arccos \left( \frac{-8}{\sqrt{\left( 4+9g^2\right) ^3}}%
\right).
\]
The states in this case all takes the form
\begin{equation}
\left| {}\right\rangle \varpropto \left( \
\begin{array}{l}
c_m\left| m\right\rangle +c_{m+2}\left| m+2\right\rangle \\
c_{m+1}\left| m+1\right\rangle
\end{array}
\right).
\end{equation}

\subsubsection{Unified expressions}

For the future use, we will give the unified expression of the eigenvalues
and eigenfucntions, which are corresponding to those in the RWA one by one
in the following.

Set $m=n-1$ in II B(1) and $m=n$ in II B(2), the eigenvalues in Eqs.
(\ref{E1nn})  and (\ref{E2nn}) can be expanded in terms of $g$ as
\begin{eqnarray}
E_{1n} &=&n+\frac 12-g\sqrt{n+1}  \nonumber \\
&&+\frac n4g^2+\frac 1{32}\frac{\left( 3n+4\right) n}{\sqrt{n+1}}g^3+...
\label{E1n}
\end{eqnarray}
\begin{eqnarray}
E_{2n} &=&n+\frac 12+g\sqrt{n+1}  \nonumber \\
&&-\frac{\left( n+2\right) }4g^2-\frac 1{32}\frac{\left( n+2\right) \left(
3n+2\right) }{\sqrt{n+1}}g^3+...  \label{E2n}
\end{eqnarray}
The corresponding eigenstates take the form
\begin{eqnarray}
\left| 1n\right\rangle &\varpropto &\left( \
\begin{array}{l}
c_n\left| n\right\rangle \\
\left( c_{n-1}\left| n-1\right\rangle +c_{n+1}\left| n+1\right\rangle \right)
\end{array}
\right),  \label{state1n} \\
\left| 2n\right\rangle &\varpropto &\left( \
\begin{array}{l}
c_n\left| n\right\rangle +c_{n+2}\left| n+2\right\rangle \\
c_{n+1}\left| n+1\right\rangle
\end{array}
\right).  \label{state2n}
\end{eqnarray}

It should be pointed out that the GS state and the first excited state can not be brought into the above
general expression for $n=0$. The GS energy and the GS state with even
parity are
\begin{equation}
E_{GS}=-\frac 12-\frac 12g^2-\frac 18g^4+...  \label{EGS}
\end{equation}
\begin{equation}
\left| GS\right\rangle \varpropto \left( \
\begin{array}{l}
c_1\left| 1\right\rangle \\
\left( c_0\left| 0\right\rangle +c_2\left| 2\right\rangle \right)
\end{array}
\right),  \label{stateGS}
\end{equation}
and the first excited state with odd parity are
\begin{equation}
E_1=\frac 12-g-\frac 12g^2+\frac 18g^3+...  \label{E1EX}
\end{equation}
\begin{equation}
\left| 1EX\right\rangle \varpropto \left( \
\begin{array}{l}
c_0\left| 0\right\rangle +c_2\left| 2\right\rangle \\
c_1\left| 1\right\rangle
\end{array}
\right).  \label{state1ex}
\end{equation}

In any case, the ratios of coefficients in the second approximation
are
\begin{equation}
c_m:c_{m+1}:c_{m+2}=\left[ -\frac{\sqrt{m+1}}{\Omega _m(E)}\right]
:1:\left[ -\frac{\sqrt{m+2}}{\Omega _{m+2}(E)}\right],
\end{equation}
where the sign $\mp \;$in Eq. (\ref{kernel}) for $\Omega _m(E)\;$ for any
eigenstates is only parity dependent.

It is interesting to note that the first two terms in Eqs.
(\ref{E1n}), (\ref {E2n}) are no other than the RWA results by Eqs.
(\ref{E1nRWA}) and (\ref{E2nRWA}) at resonance. The additional terms
appear just because of the presence of the CRTs. Also the
eigenfunctions in Eqs. (\ref{state1n}) and (\ref{state2n}) contain the
components of the RWA ones in Eq. (\ref{SRWA}). The other bare state
which can not be generated by the rotating-wave terms also emerges.
This is just our answer to the question presented in the title of
this paper.

The analytical results for the energy levels in the second order
approximation are collected in Fig. \ref{energy}(b) with red lines.
It is shown that for $g \le 0.4$, the present second order
approximation can give reasonable good results. More over, it should
be deeply impressed that analytical expressions are almost exact for
remarkable wide coupling regime $g\le$ 0.2. So it could become a
solid and concise platform to discuss the effect of CRTs on various
physical phenomena in the present experimentally accessible systems.
Note that the present maximum value for the coupling strength in the
superconducting qubit coupled to a circuit resonant \cite{Niemczyk}
has reached  $g=0.12$, to our knowledge. The application to the
Vacuum Rabi splitting is performed in the next section as a first
example.

\section{Vacuum Rabi splicings}

If we pump the dressed atom from its ground to an excited state, it
will decay to the dressed ground state through spontaneous
emissions. Under the RWA, when the atom is excited by the operator,
$V=|e\rangle \langle g|+|g\rangle \langle e| $, from the ground
state $|g,0>$ , the emission spectrum has two peaks with equal
height.  The distance of the two peaks, $2g$, is just the vacuum
Rabi splitting ~\cite {agarwal,Thompson}.

Without the RWA, we have two choices for the initial states. When
the CRTs are included, the photon in the ground state are no longer
a vacuum state, as shown in Eq. (\ref{stateGS}). In the framework of
the second order approximation, we first use $V$ to excite the atom
from the ground state at resonance
\[
\left| VRS\right\rangle ^{(1)}=V\left| GS\right\rangle \varpropto \left( \
\begin{array}{l}
c_0\left| 0\right\rangle +c_2\left| 2\right\rangle \\
c_1\left| 1\right\rangle
\end{array}
\right),
\]
which can be expanded in terms of the eigenstates with odd parity.
Note that only the following 4 excited states are included
\begin{equation}
\left| VRS\right\rangle ^{(1)}=v_1\left| 1EX\right\rangle +v_2\left|
2EX\right\rangle +v_5\left| 5EX\right\rangle +v_6\left|
6EX\right\rangle.
\end{equation}
The probabilities $h_i=\left| \nu _i\right| ^2$ can be expressed in
$g\;$as
\begin{eqnarray*}
h_1 &=&\frac 12+\frac 14g+\frac{11}{32}g^3-\frac 1{16}g^4+... \\
h_2 &=&\frac 12-\frac 14g-\frac{11}{32}g^3-\frac 1{16}g^4+... \\
h_3 &=&\frac 3{64}g^6+... \\
h_4 &=&\frac 1{16}g^4+\frac{\sqrt{3}}{48}g^5+...
\end{eqnarray*}
The atom will decay from the initial state to the dressed ground
state with an emission spectrum. The heights of the peaks in the
spectrum are proportional to the square of the probability of the
corresponding eigenstates. Therefore  one may find two main peaks,
and the other two peaks are too small to be visible, by the above
$4$ probabilities.

\begin{figure}[tbp]
\includegraphics[width=8cm]{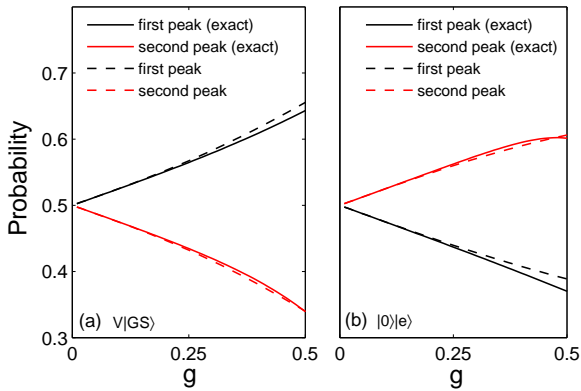}
\caption{(Color online) The peak heights (dashed lines) in emission
spectrum as a function of coupling constant at resonance for the
initial states (a) $V\left| GS\right\rangle ^{\prime }$ and
(b)$|e,0>$. The exact results are marked by black line, the present
analytical results by dashed lines. } \label{frequency}
\end{figure}
In Fig. \ref{frequency} (a), we plot the peak heights from the first
two excited states  $h_1$ and $h_2$   with dashed lines. The exact
numerical results for the heights are also list with the solid
lines. It is interesting to note that the present analytical results
for the main peak height agree excellently with the exact ones in a
wide coupling regime ($ 0<g<0.2$).

With the increase of the coupling strength, the third peak from the
sixth excited state ($h_4$) becomes visible. Recent full numerically
exact study \cite{Zhang} showed  three peaks (not four peaks) at
$g=0.8$ in their Fig. \ref{frequency}(b). Our analytical results are
consistent with this exact study qualitatively. If the third peak is
visible, the coupling constant should exceed $0.5$, the present
second order approximation can only describe it qualitatively.

The other initial states is usual one $|e,0> $ which  only include
the first and second excited states in the framework of the
second-order approximation
\begin{equation}
\left| VRS\right\rangle ^{(2)}=|e,0>=v_1\left| 1EX\right\rangle
+v_2\left| 2EX\right\rangle.
\end{equation}
We can derive the two peaks up to $o(g^6)$
\begin{eqnarray*}
h_1 &=&\frac 12-\frac 14g+\frac 5{32}g^3-\frac 1{16}g^4+... \\
h_2 &=&\frac 12+\frac 14g-\frac 5{32}g^3-\frac 1{16}g^4+...
\end{eqnarray*}
Those peak heights are also list in Fig. \ref{frequency}(b) with red
lines. It is also shown that the analytical results in this case is
consistent perfectly with the exact ones  in a wide coupling regime
($0<g<0.2$).

In both initial states, the level difference for the first two excited
states will give the vacuum Rabi splitting
\[
E_{2EX}-E_{1EX}=2g-\frac 14g^3+O(g^4).
\]
which is smaller than the RWA one by a small amount $\frac 14g^3$

For the recent experimentally accessible ultra-strong coupling
constant $g=0.1$, the effect of CRTs  on the vacuum Rabi splitting
only results in a tiny shift around $ 0.0025(\omega)$, which is too
small to be distinguished in the experimental data. While the ratios
of the two heights $h_{1}/h_{2}$ for the first and the second
initial states can be evaluated as $1.107 $ and $0.905$
respectively, which are however large enough to be seen
experimentally.

Finally, we would like to  give some remarks. As shown in Fig.
\ref{energy},  in a wide coupling regime ($ 0<g<0.2$), the
difference between the RWA energy and the present second-order
approximate energy is very subtle, but the states are essentially
different. Some bare states in the latter are absent in the former.
This is also the reason that the difference in vacuum Rabi splitting
is invisible, but in the peak heights is evident. In the JC system,
the accuracy of the eigenenergy is easy to ensure within various
approaches, but it is not so crucial, in our opinion. The components in
the eigenstates are very important, and  play the dominate role in
many physics processes.

\section{summary}

In this paper, the JC model without the RWA is  mapped to a
polynomial equation with a single variable, the eigenvalue, by the
bosonic Fock space and parity symmetry. The solutions to this
polynomial equation recover exactly all eigenvalues and
eigenfunctions of the model for all coupling strengths and
detunings. Furthermore, the analytical results are presented at
different stages. The first approximation in the present formalism
reproduces exactly the RWA results. The effect of the CRT emerges
clearly just in the second order approximation. All eigenvalues and
eigenfunctions are derived analytically. It is shown that they play
dominant role up to the remarkable  coupling strength $g=0.2$,
suggesting that they could be convincingly applied to recent circuit
quantum electrodynamic systems operating in the ultra-strong
coupling regime up to $ g=0.12$. Applications of analytical results
to the vacuum Rabi splitting are performed. Different heights of the
two main peaks are given explicitly, which agree well with the exact
ones in a wide coupling regime. The concise analytical results only
including three bare states will be very useful for the exploration
of effects of CRTs on various phenomena in the ultra-strong coupling
regime.

\section{Acknowledgement}

This work was supported by National Natural Science Foundation of
China under Grant No. 11174254, National Basic
Research Program of China (Grants No. 2011CBA00103 and No.
2009CB929104), and the Fundamental Research Funds for the Central Universities.
\appendix

\section{Solutions to univariate cubic equation}

The univariate cubic equation can be generally expressed as
\begin{equation}
x^3+bx^2+cx+d=0.  \nonumber
\end{equation}
Its solutions can be found in any Mathematics manual. If
\[
\Gamma =B^2-4AC<0,
\]
with
\[
A=b^2-3c,B=bc-9d,C=c^2-3bd,
\]
there are three different real roots with $y_1<y_2<y_3$
\begin{eqnarray}
y_1 &=&\frac{-b+\sqrt{A}\left[ \cos \theta +\sqrt{3}\sin \theta \right] }3, \\
y_2 &=&\frac{-b+\sqrt{A}\left[ \cos \theta -\sqrt{3}\sin \theta \right] }3, \\
y_3 &=&\frac{-b-2\sqrt{A}\cos \theta }3,
\end{eqnarray}
with
\begin{equation}
\theta =\frac 13\arccos \left( \frac{2Ab-3B}{2\sqrt{A^3}}\right).
\end{equation}

\end{document}